\documentclass[doublecol, longbibliography]{epl2} 
\usepackage{amsmath}
\usepackage{textgreek}
\begin{document}

\title{Spontaneous generation, enhanced propagation and optical imprinting of quantized vortices and dark solitons in a polariton superfluid: towards the control of quantum turbulence}
\shorttitle{Controlled Quantum Turbulence in polariton superfluids} 

\author{A. Ma\^{i}tre\inst{1}\and F. Claude\inst{1}\and G. Lerario\inst{1} \and S. Koniakhin\inst{2} \and S. Pigeon\inst{1}\and D. Solnyshkov\inst{2}\and G. Malpuech\inst{2}\and Q. Glorieux\inst{1}\and E. Giacobino\inst{1}\and A. Bramati\inst{1}}
\shortauthor{A. Ma\^{i}tre \etal}
\institute{                    
  \inst{1} Laboratoire Kastler Brossel, Sorbonne Universit\'{e}, CNRS, ENS-Universit\'{e} PSL, Coll\`{e}ge de France - 75005 Paris, France\\
  \inst{2} Institut Pascal, PHOTON-N2, Université Clermont Auvergne, CNRS, SIGMA - Clermont, F-63000 Clermont-Ferrand, France}

\abstract{In resonantly pumped polariton superfluids we recently explored a new regime based on the bistability of the polariton system to enhance the propagation of polariton fluids up to macroscopic distances. This technique together with an all-optical imprinting method allowed the generation and control of various topological excitations such as quantized vortices and dark solitons. The flexibility and scalability of the new experimental scheme opens the way to the systematic study of quantum turbulence in driven dissipative quantum fluids of light. In this article we review the basic working principles of the bistability enhanced propagation and of the imprinting technique and we discuss the main achieved results as well as the  most promising future research directions.}


\maketitle

\section{Introduction}
Exciton-polaritons are half-light half-matter quasi-particles coming from the strong coupling between excitons and photons in semiconductor microcavities\cite{Kavokin2007a}. They inherit specific properties from their components: a very light mass coming from the photon component and strong mutual interactions from their excitonic nature. 

In the last decade these systems have demonstrated to be an ideal playground for the study of out-of-equilibrium condensates and 2D quantum fluid hydrodynamics \cite{Carusotto2013a}. 

In particular, the creation of polariton fluids via an all optical excitation enabling a full control of the speed of polariton flows allowed the hydrodynamic generation of a rich variety of topological excitations in such systems, ranging from quantized vortices to  dark solitons, via the interaction of a supersonic polariton wavepacket with a structural defect\cite{Amo2009, Amo2009a, Grosso2011, Amo2011, Roumpos2011, Nardin2011, Lerario2017a}

Pulsed resonant excitation as well as continuous-wave (cw) resonant pumping have been used. However, both the configurations exhibit the same fundamental limitation: the polariton density strongly decays along the propagation due to the short polariton lifetime. As a result the propagation distances accessible in these early experiments are quite short, strongly limiting the study of the dynamics of the topological excitations.

In a series of recent articles, devoted to the deep study of the properties of the bistability exhibited by the polariton systems under resonant pumping, we investigated a new configuration and demonstrated that in the bistable regime it is possible to get rid of the polariton density decay generating a superfluid flow propagating for macroscopic distances, typically one order of magnitude longer than the previous observations.

Remarkably, in the bistable regime the topological excitations can be generated and their propagation sustained and strongly enhanced far beyond the polariton free propagation length. 

Moreover, we implemented an all optical imprinting technique which allowed us to generate in a fully controlled way dark soliton pairs in different regimes and to study their stability against the onset of the snake instabilities. The observation of the breaking of dark solitons in vortex streets illustrates the high potential of this method for the systematic study of the quantum turbulence in polariton quantum fluids.

\section{Theoretical model}
The standard way to describe the dynamics of a resonantly driven polariton fluid is to use a generalized Gross Pitaevskii equation. In the exciton-photon basis, the system is  described by the following coupled equations: 

\begin{multline}
\label{eq:GPEXcav}
    i \hbar \dfrac{d}{dt} 
    \begin{pmatrix}
    \Psi_{\gamma}
    \\ \Psi_{X}
    \end{pmatrix} = 
    \begin{pmatrix}
    \hbar F_{p}
    \\ 0
    \end{pmatrix} + 
    \Bigg[ H_{lin} 
    + \\
    \begin{pmatrix}
    V_{\gamma}
    - i \hbar \gamma_{cav} & 0 \\
    0 & V_{X}
    - i \hbar \gamma_{X}
    + \hbar g n
    \end{pmatrix}
    \Bigg]
    \begin{pmatrix}
    \Psi_{\gamma}
    \\ \Psi_{X}
    \end{pmatrix}
\end{multline}
where the fields $\Psi_{X}
$ and $\Psi_{\gamma}
$, describe the excitons and photons, respectively.

The system losses are represented by the two linewidths $ \gamma_{X}$ and $ \gamma_{cav}$.
The continuous pumping $F_{p}
$ applies to the photonic field while the interactions defined by $g$ occur between excitons; $n$ is the excitonic density. 
The external potentials $V_{\gamma}
$ and $V_{X}
$ come from the photonic and excitonic defects naturally present in the microcavity.
$H_{lin}
$ is the linear Hamiltonian: 

\begin{equation}
    H_{lin}
    = 
    \begin{pmatrix}
    E_{X}^{0} & \dfrac{\hbar}{2} \Omega_{R} \\
    \dfrac{\hbar}{2} \Omega_{R} & 
    E_{\gamma}(-i \nabla_{\mathbf{r}})
    \end{pmatrix}
\end{equation}

The equation (\ref{eq:GPEXcav}) can also be written in the polariton basis, and in particular if we focus on the lower polariton branch, we obtain the driven-dissipative Gross-Pitaevskii equation for the polariton field:

\begin{equation}
\label{eq:ddGPE}
i\hbar \dfrac{\partial}{\partial t} \Psi
=\bigg(-\dfrac{\hbar^{2}}{2m_{LP}^{*}} \nabla_{\mathbf{r}}^{2} + V
- i \hbar \gamma
+ \hbar g_{LP} n
\bigg) \Psi
+ \hbar F_{p}
\end{equation}

with $m_{LP}^{*}$ the effective mass of the lower polariton, $V = |X_{\mathbf{k}}|^{2} V_{X} + |C_{\mathbf{k}}|^{2} V_{\gamma}$ the external potential felt by the lower polaritons and $g_{LP} = |X_{\mathbf{k}}|^{4} g$ the interaction constant between the lower polaritons in the same modes.

As we are interested only in the lower polariton branch, to lighten the notation, the indices will be removed: $g_{LP} = g$ and $m_{LP}^{*} = m^{*}$.

This equation is very similar to the standard Gross-Pitaevskii which describes the atomic BECs, except for the loss term $-i \hbar \gamma$ and the pump term $F_{p}(\mathbf{r}, t)$.
It is indeed an important difference between our system and cold atom gases: the polariton lifetime $\tau = \dfrac{\hbar}{2 \pi \gamma}$ is of the order of some tens of picosecond, compared to seconds for atoms. 
Hence the fact that the losses need to be continuously compensated: our system is out of equilibrium.

Let us now focus on the mean-field stationary solutions of equation (\ref{eq:ddGPE}) in the homogeneous case, \textit{i.e.} for an external potential equal to zero.
The system is driven by the pump field $F_{p}(\mathbf{r}, t) = F_{p}(\mathbf{r}) e^{i(\mathbf{k}_{p}\mathbf{r} - \omega_{p}t)}$.
Therefore, the solutions can be written as $\Psi(\mathbf{r}, t) = \Psi_{0}(\mathbf{r}) e^{i(\mathbf{k}_{p} \mathbf{r} - \omega_{p}t)}$.
It results in the mean field stationary equation:

\begin{equation}
\label{eq:statsolddGPE}
    \bigg( - \dfrac{\hbar^{2} \mathbf{k}_{p}^{2}}{2m^{*}}
    - i \hbar \gamma 
    + \hbar g n_{0}(\mathbf{r}) \bigg)
    \Psi_{0}(\mathbf{r}) + \hbar F_{p}(\mathbf{r}) = 0
\end{equation}

This equation is responsible for the bistability phenomenon observed in polariton system and induced by a quasi-resonant pumping.
Indeed, the crucial parameter which determines the bistable behaviour is the detuning between the pump and the lower polariton branch $\Delta E_{lasLP} = \hbar \omega_{p} - \hbar \omega_{LP}$.
Multiplying the stationary equation by its complex conjugate, we obtain the equation for the intensity:

\begin{equation}
\label{eq:bistabilityIntensity}
    I(\mathbf{r}) = \bigg( (\hbar \gamma)^{2} 
    + \Big( \Delta E_{lasLP} - \hbar g n_{0}(\mathbf{r}) \Big)^{2}
    \bigg) n_{0}(\mathbf{r})
\end{equation}

with $I(\mathbf{r})$ the intracavity intensity.
By now this equation can be derived into:

\begin{equation}
    \dfrac{\partial I}{\partial n_{0}} = 3 g^{2} n_{0}^{2}
    - 4 \Delta E_{lasLP} g n_{0} + (\hbar \gamma)^{2}
    + \Delta E_{lasLP}^{2}
\end{equation}

the discriminant of which is $(2g)^{2} ( \Delta E_{lasLP}^{2} - 3 ( \hbar \gamma)^{2} ) $. Therefore, the previous equation can have two distinct roots if the detuning satisfies the condition $\Delta E_{lasLP} > \sqrt{3} \hbar \gamma$.

\begin{figure}[h]
    \centering
    \includegraphics[width=0.95\linewidth]{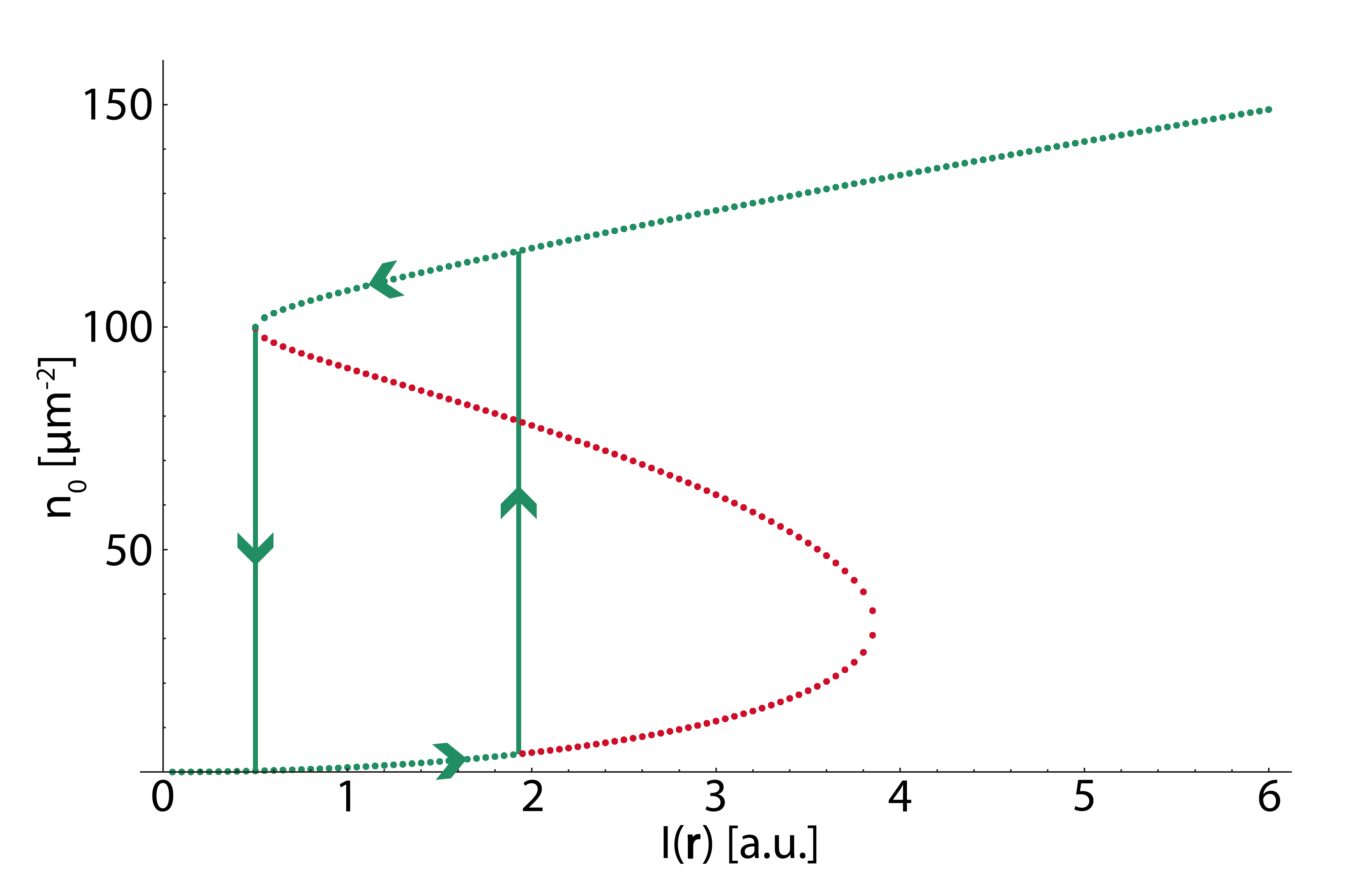}
    \caption{Numerical resolution of equation \ref{eq:bistabilityIntensity} for $\hbar \gamma = 0.1$ meV and $\Delta E_{lasLP} = 1$ meV. The green region shows the stable solutions while the red one indicates the unstable ones \cite{Ciuti2005}. The density jumps from one branch to the other at the frontier between the green and red regions: we have a hysteresis cycle illustrated by the arrows.}
    \label{fig:BistThom}
\end{figure}

In this case, the system presents a bistable behaviour: the polariton density $n_{0}(I)$ displays a range of intensities for which the system has two possible outputs, as plotted in figure \ref{fig:BistThom}.
The system is stable on the green parts of the curve, while the red region shows non-physical solutions \cite{Ciuti2005}. 
The system therefore jumps from one branch to another at the limits between the red and green regions. 
We observe a bistability cycle: within the bistable range, the upper branch is only accessible by lowering the pump intensity from a higher one, while the lower branch can only be reached by increasing the intensity from a lower one, as indicated by the arrows.


A very specific behaviour of the system when it is excited in the bistable range was pointed out in \cite{Pigeon2017}: the authors suggested to use the optical bistability of an exciton-polariton system to enhance the propagation of the polariton superfluid. 
In particular, the use of two different beams, simultaneously exciting the microcavity, can enable the generation, over a macrocopic scale, of a high density, bistable, polariton fluid.

\begin{figure}[h]
    \centering
    \includegraphics[width=0.95\linewidth]{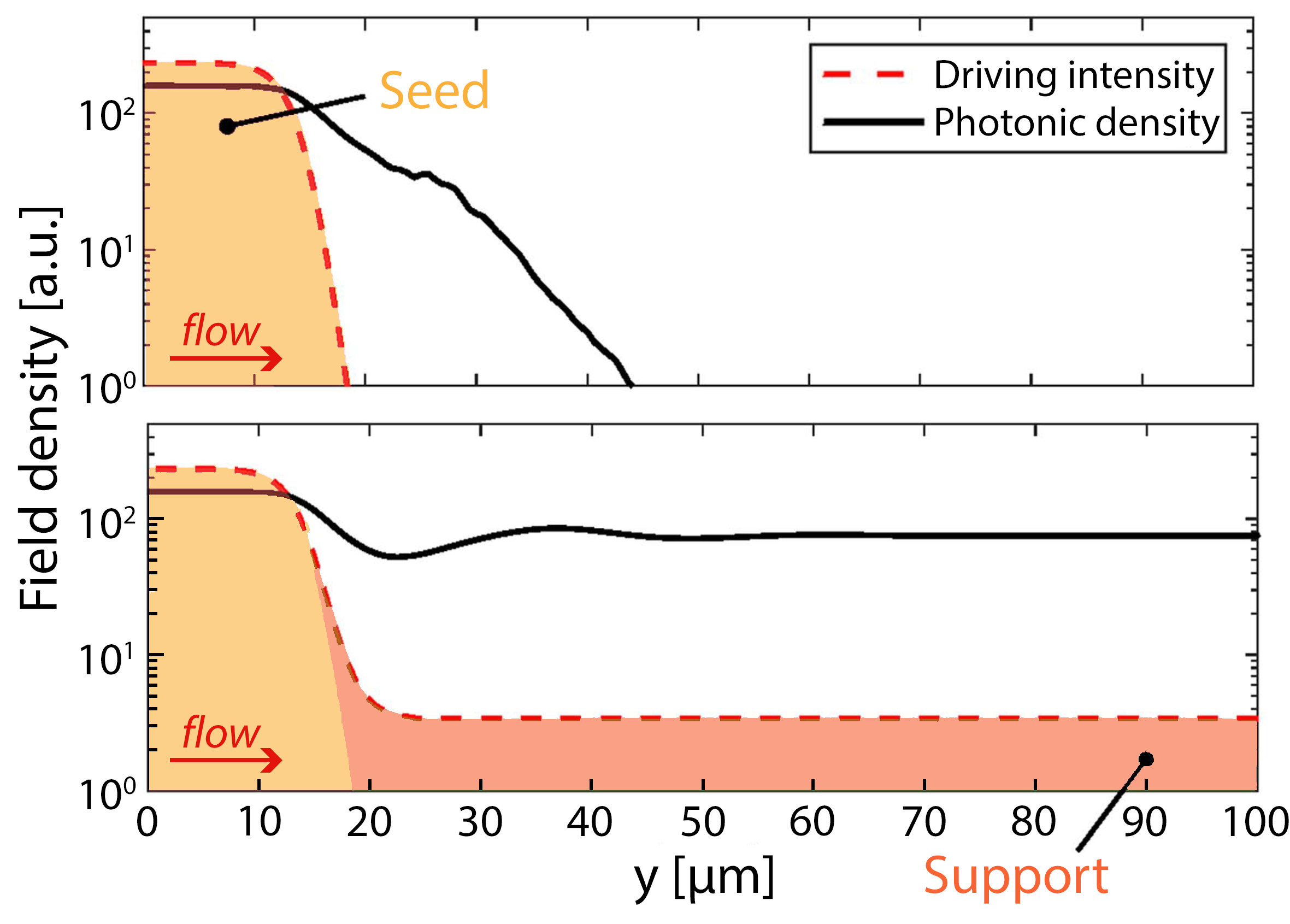}
    \caption{The black solid line illustrates the intracavity intensity (logarithmic scale), while the red dashed one is the driving intensity sent to the system. The colored regions delimits the different driving fields: the seed in yellow and the support in orange. The polariton flow goes from left to right. On the upper panel, only the seed is sent: it creates a high density polariton fluid, decaying exponentially out of the pumping region. The propagation length is limited by the polariton lifetime. In the lower panel, the support field is added: even if its density is two order of magnitude lower than the seed one, the total polariton density is maintained at its higher level all along the presence of the support.
    From \cite{Pigeon2017}}
    \label{fig:SeedSuppDens}
\end{figure}

Let consider two driving fields, with the same frequency $\omega_{p}$ and the same in-plane wavevector $\mathbf{k}_{p}$.
The first one, called the seed, is localized in space and has a high intensity: $I_{r} > I_{high}$. It thus produces a nonlinear superfluid above the bistability cycle.
The second field is the support, ideally an infinitely extended constant field, stationary in time and homogeneous in space. Its intensity is weaker than the intensity of the seed one, and is placed inside the bistability cycle: $I_{low} < I_{s} < I_{high}$.
The goal of this configuration is to enhance the propagation and density of the polariton fluid by combining the properties of both beams. Numerical simulations were done to understand their combination, presented in figure \ref{fig:SeedSuppDens}. 
They have been realized for a cavity without any defects ($V=0$) such that: $\hbar \omega_{C}(\mathbf{k} = \mathbf{0}) = 1602$ meV, $\hbar \omega_{X}^{0} = 1600$ meV, $\hbar \gamma_{X} = \hbar \gamma_{C} = 0.05$ meV, $\hbar \Omega_{R} = 2.5$ meV and $\hbar g = 0.01$ meV/\textmu m\textsuperscript{2}. 
The driving field parameters are $\delta E = \hbar \omega_{p} - \hbar \omega_{LP}(\mathbf{k} = \mathbf{k}_{p}) = 1$ meV and $ |\mathbf{k}_{p}| = (0.1)^{T} $ \textmu m\textsuperscript{-1}.

The field intensity is shown as a function of space, plotted in logarithmic scale. The flow of particles goes from left to right. The driving intensity sent in the system is pictured with the red dashed line, while the black solid line shows the steady-state photonic intracavity density.

On the upper panel, only the seed is sent on the left, illustrated with the yellow highlighted region ($F_{r}(\mathbf{x}) \neq 0$; $ F_{s} =0 $). 
Given the presence of an in-plane wavevector, the polariton density expands, but it decays exponentially due to the finite polariton lifetime.
Even with the present best quality samples, the propagation distance in this configuration is limited to around 50 microns. Moreover, as the density is decreasing all along propagation, all the related fluid parameters are also constantly changing.

\begin{figure}[h]
    \centering
    \includegraphics[width=0.9\linewidth]{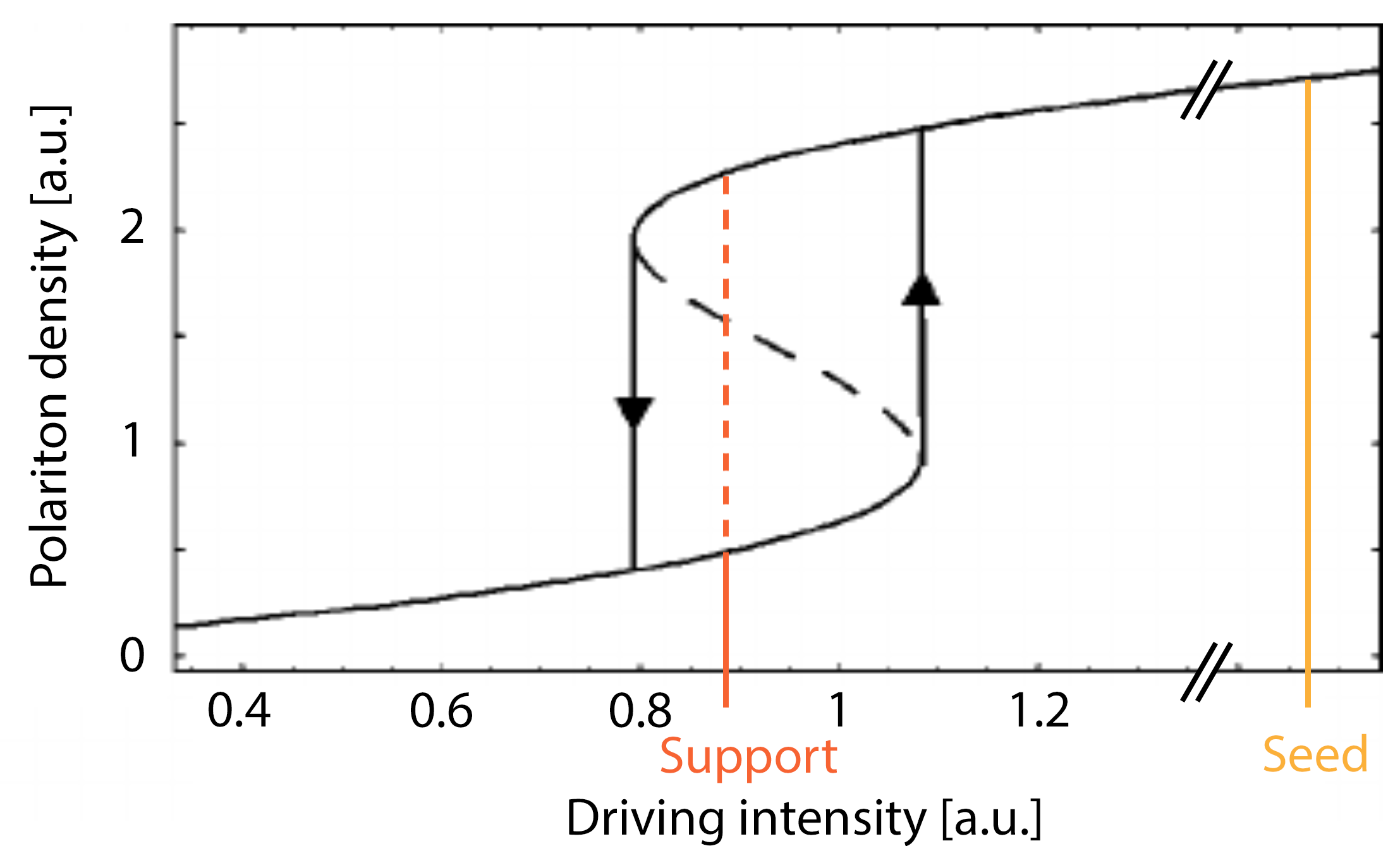}
    \caption{The seed is chosen to be above the hysteresis cycle. Only one stable state is available at this place, on the upper nonlinear branch of bistability. The support has much lower intensity, inside the bistable region. When both fields are sent, the seed presence ensures the system to stay on the upper branch, everywhere the support is present. 
    }
    \label{fig:SeedSuppBist}
\end{figure}

The lower panel shows the behaviour of the system when the support, in orange, is added to the previous configuration.
The support intensity is two order of magnitude lower than the seed one, but infinitely extended in space for this simulation.
The presence of the support field has a strong impact on the total polariton density: its high level created at the seed location is maintained without any decay all over the region where is the support, despite its much weaker intensity.
Bistability is essential to explain this behaviour. The seed is placed outside the bistability cycle, where only one stable state is accessible (see figure \ref{fig:SeedSuppBist}). It therefore ensures the system to be on the upper branch of the cycle, the nonlinear branch.
On the other hand, the support is chosen to be inside the bistability cycle, where two states are reachable. The support itself can not reach the upper branch and, if alone, would only drive the system in the linear configuration. 
Yet, as the seed places the system in the nonlinear branch and touches the support illuminated region, by extension, the neighbor region jumps also on the upper branch which expands to all the support area.

\section{Subsonic flow: vortex stream generation}

The idea of the seed-support configuration is not only to obtain an extended fluid of polaritons, but also to study its properties, and in particular its ability to generate topological excitations, such as dark solitons and quantized vortex-antivortex pairs.
The different hydrodynamic regimes of a polariton fluid and their effects on the generation of topological excitations have been previously studied in the case of direct injection \cite{boulier2015vortex,boulier2016injection} and for a single intense and localized pump, placed upstream of a structural defect \cite{Pigeon2011, Amo2011}.
The presence of the defect creates turbulence along the flow, which evolves differently depending on the ratio between the speed of the fluid and the sound velocity, \textit{i.e.} the Mach number of the system: $M = v_{f}/c_{s}$.
The speed of the fluid that needs to be taken into account is the one around the defect, therefore always higher than the one extracted far from the defect. Indeed, due to the impenetrable nature of the defect, the particles flowing close to it are accelerated \cite{Frisch1992}, which induces a phase shift as the fluid phase and speed are connected.
In the case of a global subsonic flow, but locally supersonic around the defect, vortex-antivortex pairs emerge in the wake of the defect. Increasing the Mach number induces an increase in the emission rate of the vortices, which will finally merge together in a pair of dark solitons for a Mach number close to 1 or higher.

To numerically reproduce the effect of a cavity structural defect, a large potential photonic barrier is introduced ($V \neq 0$). 
The seed is placed upstream to it, localized and with a fixed intensity above the bistability cycle.

The bistable regime ensures the release of the phase constraint imposed by a resonant driving. 
Thus, even though the support field possesses a flat phase, vortex-antivortex pairs are generated in the wake of the defect as long as the support intensity places the fluid in the bistable regime.  The seed-support configuration previously described is used to enhance their propagation length by one order of magnitude.

\begin{figure}[ht]
    \centering
    \includegraphics[width=0.7\linewidth]{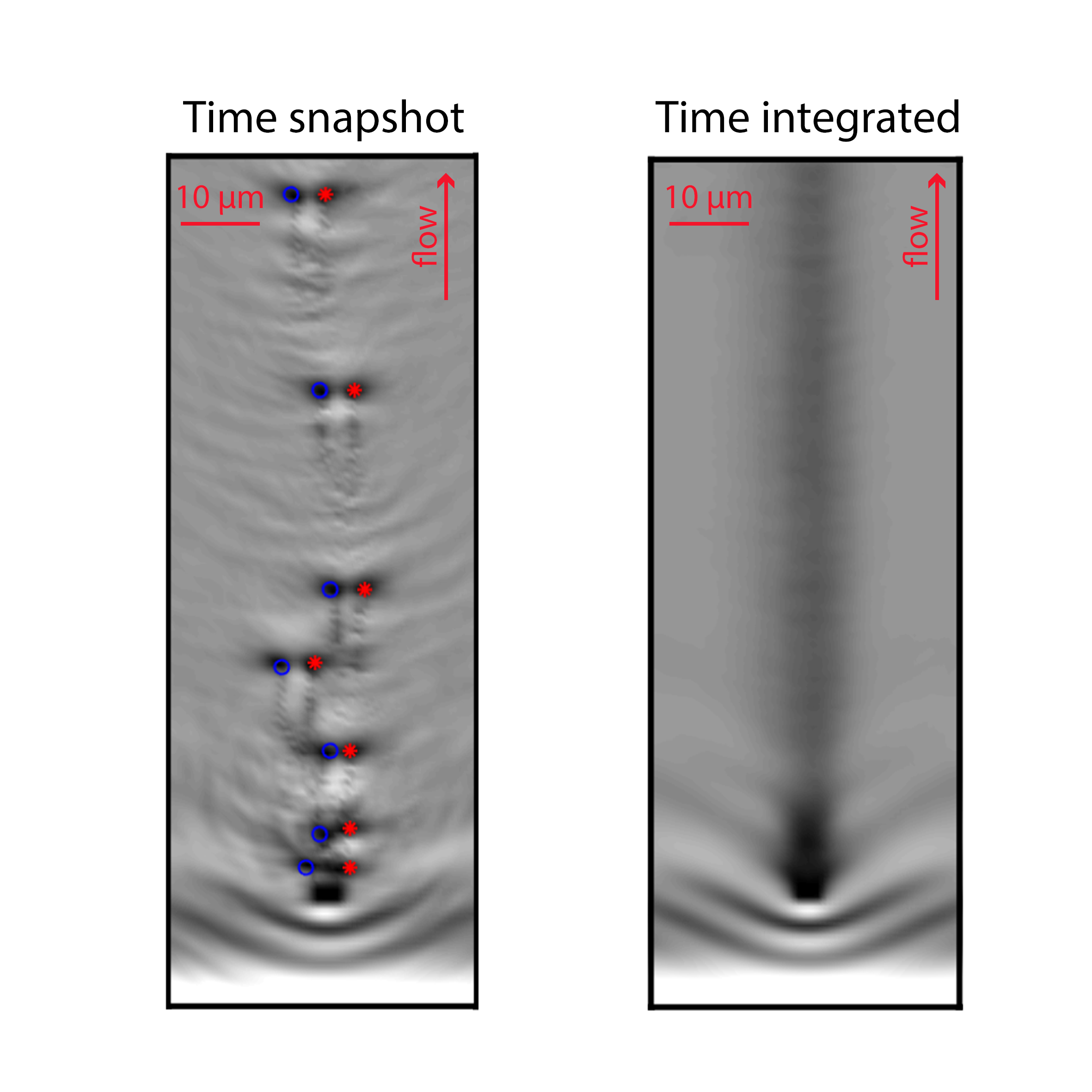}
    \caption{On the left, the time snapshot shows the position of vortex and antivortex flowing in the wake of the defect. On the right, the time integrated image has blurred the positions of the vortices and results in a thick shadow along the flow.}
    \label{fig:VortSnapInt}
\end{figure}

Time resolved numerical simulations have been made in order to precisely locate the vortex-antivotex pairs, however as the experiment is in continuous wave excitation, time integrated numerical simulations were also performed. Figure \ref{fig:VortSnapInt} shows the comparison between a snapshot image on the left, and, with the same parameters, an image with 1 ms integration time, on the right.

Vortex and antivortex have opposite circulation, and are spotted by red and blue dots, respectively.
Vortex-antivortex can stay bounded and propagate along the flow side by side, or annihilate each other and vanish.


A time average flow of vortex pairs appears as a thick line of lower density, the dip height of which is proportional to the vortex density. The phase pattern would be blurried and a decrease of visibility would appear along the vortex stream.

\begin{figure}[h]
    \centering
    \includegraphics[width=0.85\linewidth]{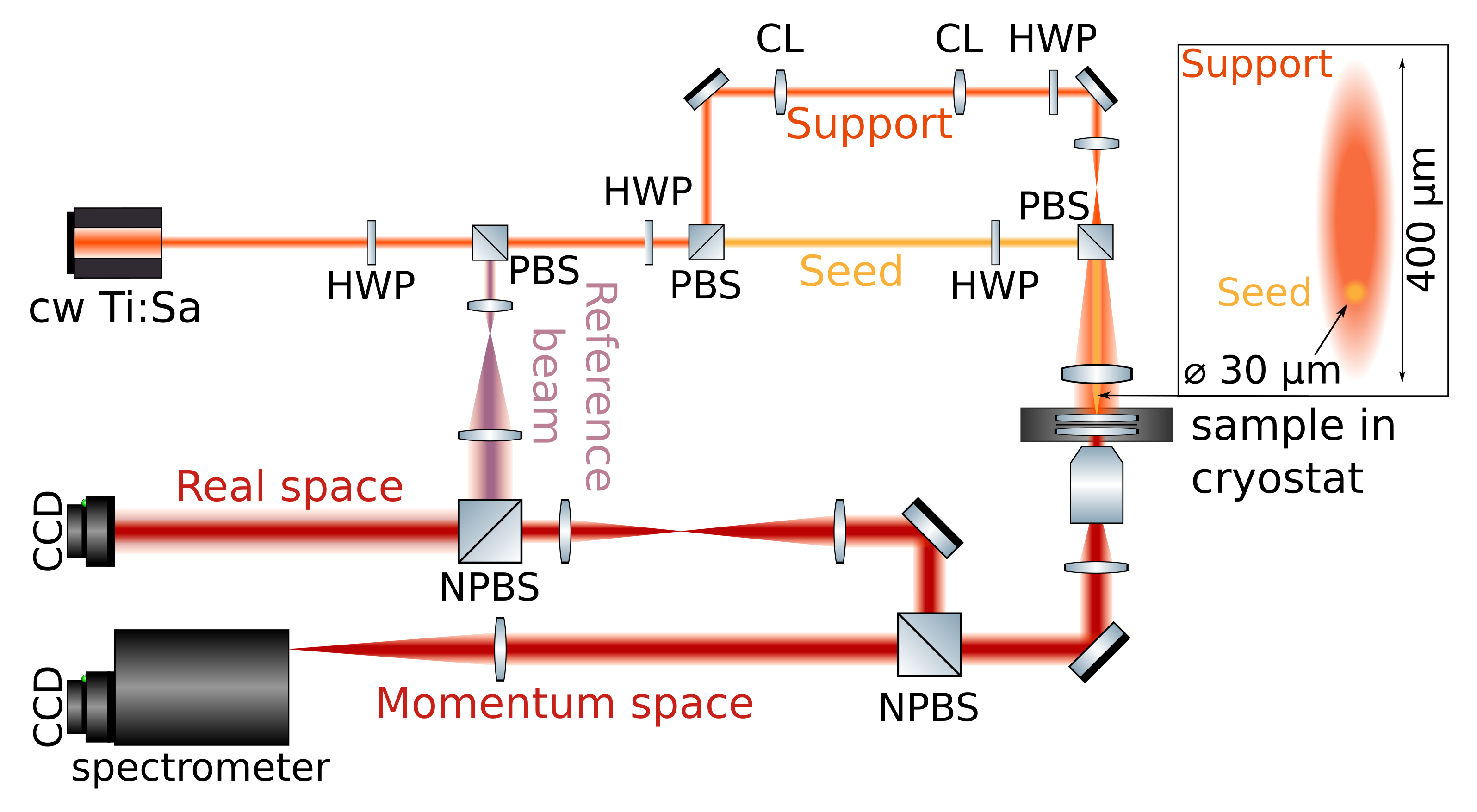}
    \caption{Experimental setup. The CW Ti:Sa laser beam is split into three using half waveplates (HWP) and polarizing beam-splitter (PBS). A reference beam, in light red, is set aside for interferogram on the detection part; the seed beam in yellow is focused on the sample to a spot of 30 microns diameter; the support in orange is extended and elongated in the vertical direction by cylindrical lenses (CL) and sent into the microcavity. The seed and the support, not centered to one another (see inset), share the same wavevector. The detection is done in real space, from which we can get information on the density and phase maps, and also in momentum space through the spectrometer.}
    \label{fig:SetupSeedSupp}
\end{figure}

The experimental setup  is displayed in figure \ref{fig:SetupSeedSupp} \cite{Lerario2020}. The initial laser source is a CW Titanium Sapphire laser. It is split up a first time into the main beam and the reference beam, later used to realize interferograms and get information on the phase. The main beam is split again to generate the seed and the support beams. The seed is then focused on the sample into a spot of 30 microns diameter, for an intensity $I_{r} = 10.6$ W/mm\textsuperscript{2}. On the other hand, the support is shaped by two cylindrical lenses in an elliptical spot of 400 microns length, elongated in the $y$ direction with an intensity of 5.8 W/mm\textsuperscript{2}. The inset of figure \ref{fig:SetupSeedSupp} gives a representation of the relative position of the beams; the seed is not centered on the support so that the topological excitations can be studied on the flat part of the Gaussian support beam.

The real space detection gives access to the polariton density in the plane of the cavity. The reference beam interacts with the signal of the cavity and the resulting interferogram is used to reconstruct the phase map. All the other parameters of the excitation are coming from the momentum space data, acquired via a spectrometer.

The sample under investigation is a GaAs/AlGaAs microcavity with 21/24 (front/back) layers of DBR and In$_{0.04}$Ga$_{0.96}$As quantum wells at each of the three antinodes of the confined electromagnetic field \cite{Houdre2000}. A small wedge is inserted between the Bragg mirrors during the fabrication process allowing to precisely select the cavity-exciton detuning.  The half Rabi splitting is 2.55~meV. The polariton mass, extracted from the dispersion,  is $7\cdot 10^{-5}$ free electron mass. The polariton lifetime is about 14 ps. The experiments are performed in a liquid helium cryostat at cryogenic temperature in transmission configuration.

\begin{figure}[h]
    \centering
    \includegraphics[width=\linewidth]{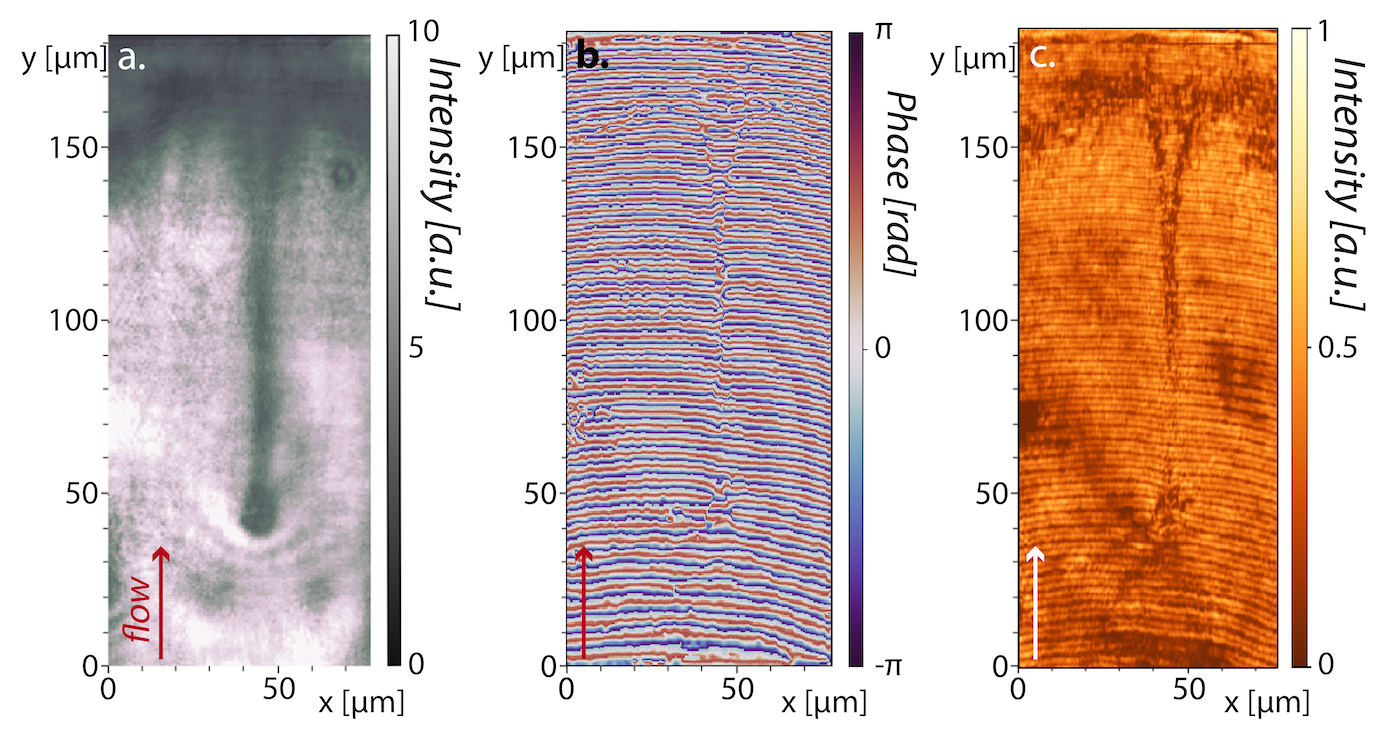}
    \caption{Vortex stream generation. a. Time integrated density map of a flow of vortex pairs generated in the wake of a defect. As the vortices move along the flow, the time integrated image results in a blurried density dip in the wake of the defect. b. Interferogram of the previous image, showing phase irregularities along the vortices propagation. c. Visibility map extracted from the interferogram, displaying a lower visibility along the vortex stream.} 
    \label{fig:VortSeedSuppIphVis}
\end{figure}

The results when both seed and support are sent together are shown in Figure \ref{fig:VortSeedSuppIphVis}.
As predicted, a shadow appears in the wake of the defect, which corresponds exactly to the time integrated image numerically calculated and presented in figure \ref{fig:VortSnapInt}. The pairs of vortex-antivortex are generated around the defect and follow the flow, leading to a decrease of density along their propagation path on the time integrated image.

To confirm that this shadow is indeed due to the presence of vortices, an interferogram has been realized as well as a visibility map, displayed on figures \ref{fig:VortSeedSuppIphVis}b. and c.
As the vortices are moving along the flow, one cannot expect to observe forks, the typical signature of vortices, due to the low time resolution of the detection (the integration time is typically of few milliseconds, while the flow speed is about 1 \textmu m/ps).
However, the presence of the vortices is visible on the phase map as a blur of the fringes along their path. 

In order to further insure that the vortex generation is responsible for the shadow, a visibility map has been extracted from the interferogram and shown in figure \ref{fig:VortSeedSuppIphVis}c. If the irregularities of the phase pattern are indeed a phase blur due to the flow of vortices, the fringes visibility should decrease. This is exactly what is observed: the shadow and the phase variations coincide with a dip in visibility, confirming its attribution to the vortex stream.

Another interesting feature visible in this figure is that at the top end of the fluid, the stream separates into two thinner and darker lines.
This is due to the fact that in this region, vortices merge together and become grey solitons. 
Indeed, the vortex core size is controlled by the hydrodynamic properties of the system, and in particular its healing length $\xi$, defined as \cite{Carusotto2013a}:

\begin{equation*}
\xi = \dfrac{\hbar}{\sqrt{2m^{*}\hbar gn}}
\end{equation*}

with $m^{*}$ the effective mass of the polariton and the product $gn$ the polariton interaction energy. 
On the top part of figure \ref{fig:VortSeedSuppIphVis} a. the density of the non linear fluid on the upper bistability branch is slowly decreasing. This density decrease has a direct impact on the healing length which inversely increases, leading to vortex cores bigger and bigger: eventually vortices merge together into a grey soliton.
A precise phase jump is indeed visible at the corresponding position of the interferogram.
It can also be explained in terms of Mach number, as the decrease of density induces a decrease of the sound speed and therefore an increase of the Mach number: the flow becomes supersonic there.

\section{Supersonic flow: parallel dark soliton pair generation}

In order to study the generation of solitons, the supersonic configuration is investigated. According to previous studies in polaritons superfluid \cite{Amo2011}, high Mach numbers allow for the generation of dark solitons. 
However, the system configuration, without the support beam, did not allow for long propagation: the quasi-resonant pump beam was sent upstream of the defect and the solitons could only be observed for short distances (around 30 micrometers), limited by the exponential decay of the polariton density due to the finite polariton lifetime. 

The goal here is to use the configuration of seed-support excitation to generate dark solitons propagating over long distances and to enable the study of their hydrodynamic behaviour.

The setup of this experiment is similar to the previous one for the vortex generation, with two co-propagating beams sent to the cavity: the seed, localized and intense, and the support, extended and with an intensity within the bistability cycle. 

\begin{figure}[h]
    \centering
    \includegraphics[width=0.6\linewidth]{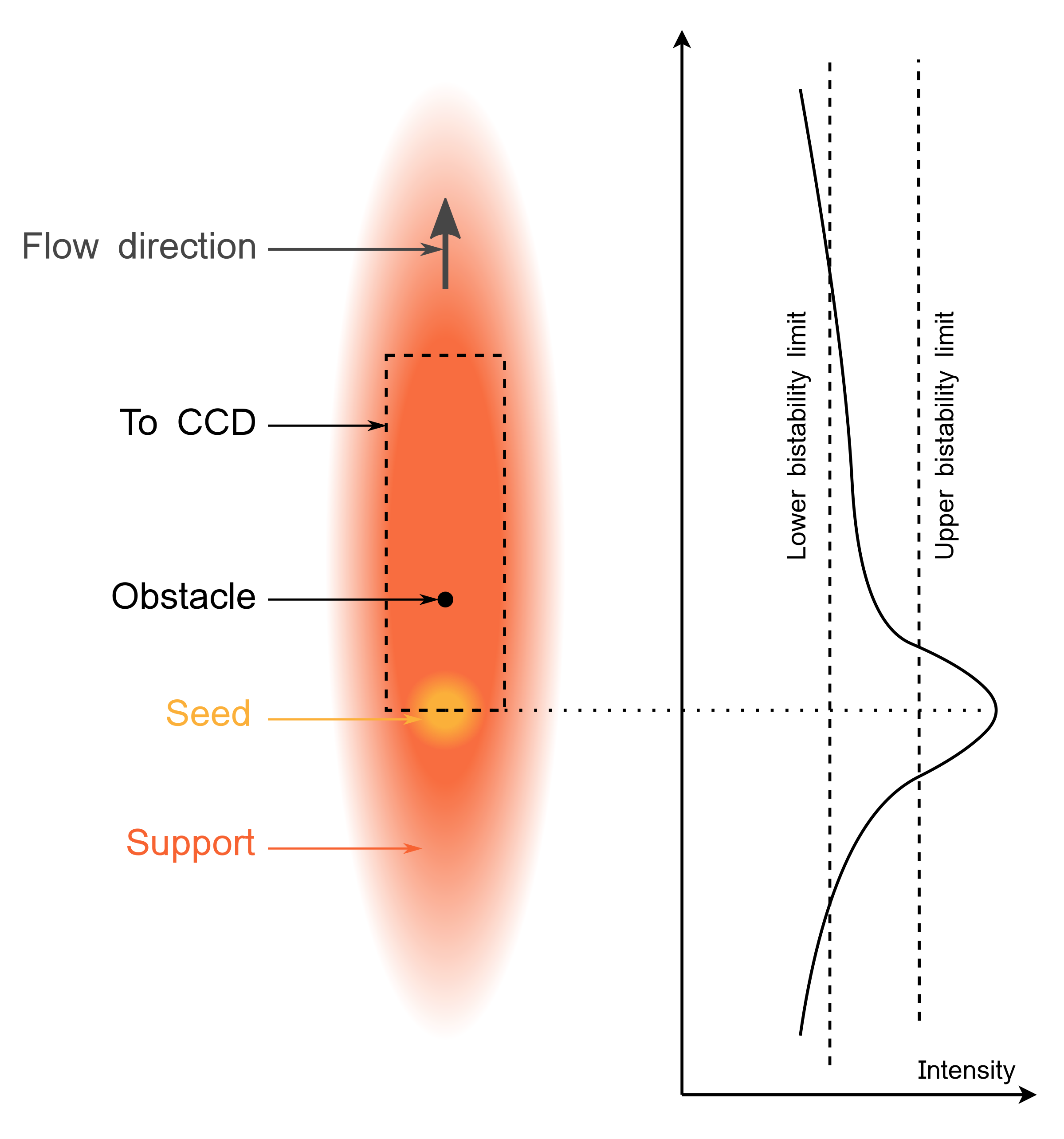}
    \caption{The support beam is elongated along the vertical direction and its intensity is within the bistability cycle. The flow is from bottom to top, and a structural defect is placed in the center of the support. Upstream to it is sent the seed, localized and with an intensity above the bistability cycle. The distribution of the total intensity along the vertical axis is displayed on the right as well as the upper and lower bistability limits.}
    \label{fig:SolBeams}
\end{figure}

The position of the seed and the support beams on the sample are shown in figure \ref{fig:SolBeams}. The seed, shifted compared to the support center, is placed upstream of the considered structural defect - the flow is from bottom to top. 
The curve on the right shows the distribution of the intensity in the cavity along the vertical axis. The region close to the seed is above the upper bistability limit, while the main part of the support is inside the cycle. The combination of the two beams ensures the fluid to be bistable and on the upper branch of the cycle.

To generate dark solitons in the wake of a defect, supersonic conditions need to be created. The in-plane wave vector is chosen to be high ($k = 1.2$ \textmu m\textsuperscript{-1}) in order to ensure a high velocity of the fluid: in this case, $v_{f} = 1.52$ \textmu m/ps. 
The speed of sound is extracted through the energy renormalization and is measured to be $c_{s} = 0.4$ \textmu m/ps: the supersonic conditions are reached.

\begin{figure}[h]
    \centering
    \includegraphics[width=0.6\linewidth]{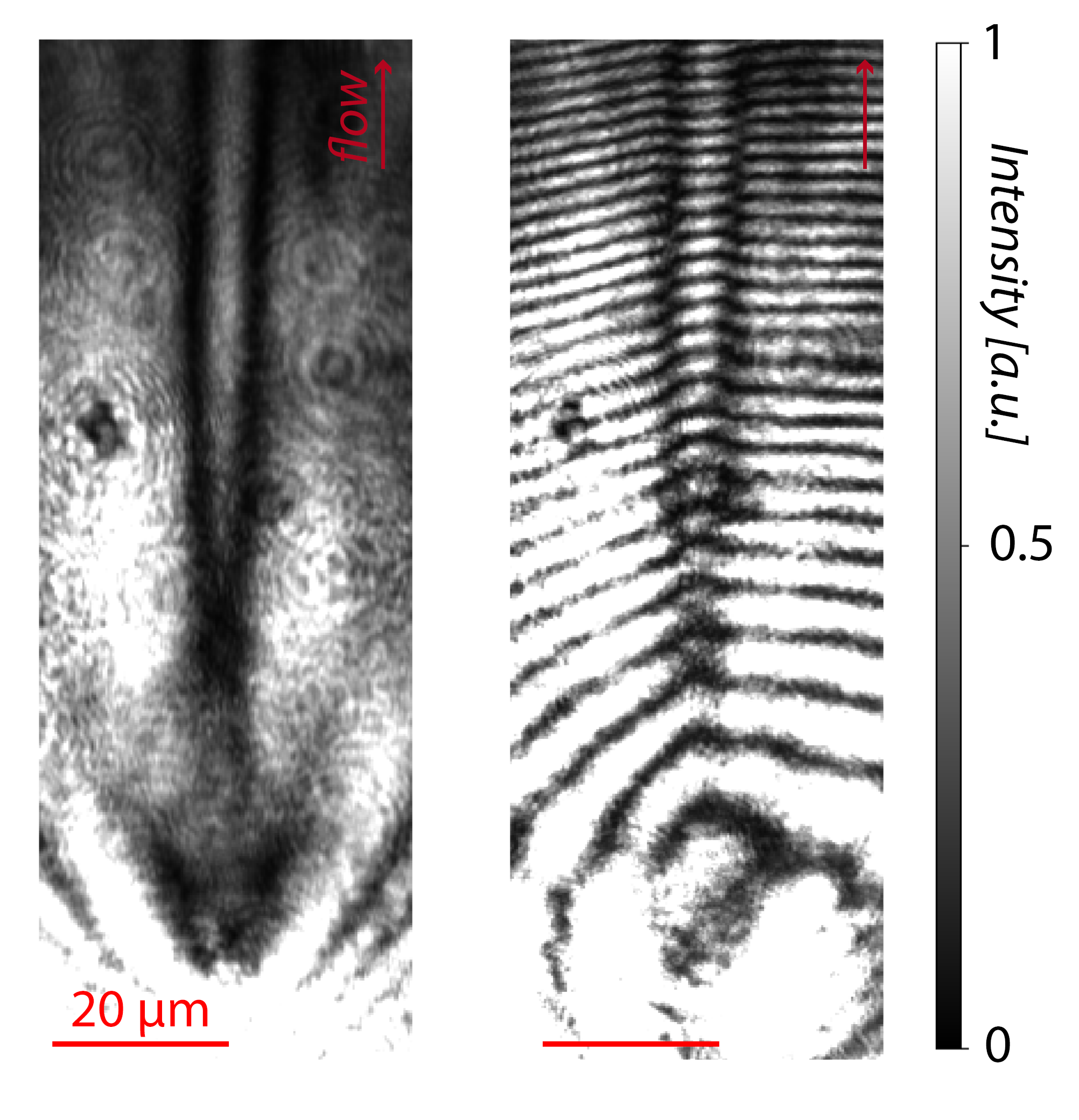}
    \caption{Spontaneous generation of a pair of parallel dark solitons. Intensity (left) and interferogram (right) of a dark soliton pair spontaneously generated in the wake of a structural defect. The solitons propagate after a short distance align and propagate parallel for over a hundred microns.}
    \label{fig:SolParrExp}
\end{figure}

The figure \ref{fig:SolParrExp} presents the observation of a spontaneous generation of dark solitons in the wake of a defect: intensity on the left and interferogram on the right.
The first conclusion that can be done from those results is that the propagation length is indeed greatly enhanced. The scale bar illustrates 20 microns: the solitons are sustained for more than hundred microns, one order of magnitude more than previously reported.

Moreover the solitons have a surprising behaviour. They are generated at the same position, close to the defect, and propagate away from each other for a few microns; eventually, they reach an equilibrium separation distance of about 8 microns, align and stay parallel as long as they are sustained. The observation of such a bound state of dark solitons is quite unexpected as solitons have repulsive interactions and usually constantly repel each other \cite{Zakharov1973}.

The observed solitons are fully dark: the intensity dip goes to zero and the phase jump across the solitons is nearly \textpi\ all along the propagation, confirming their transverse velocity is close to zero.

To fully understand the phenomenon taking place in the system, numerical simulations based on the coupled equations of the excitons and cavity photons fields, $\psi_{X}$ and $\psi_{\gamma}$  were realized to reproduce the solitons behaviour.

The presence of a structural defect is modeled in the equations, by the term $V(\mathbf{r})$, representing a 10 meV potential barrier, with a Gaussian shape of 10 \textmu m width. 
The spatial profiles of the seed and support are also accurately modeled through the pumping term. 

\begin{figure}[h]
    \centering
    \includegraphics[width=0.6\linewidth]{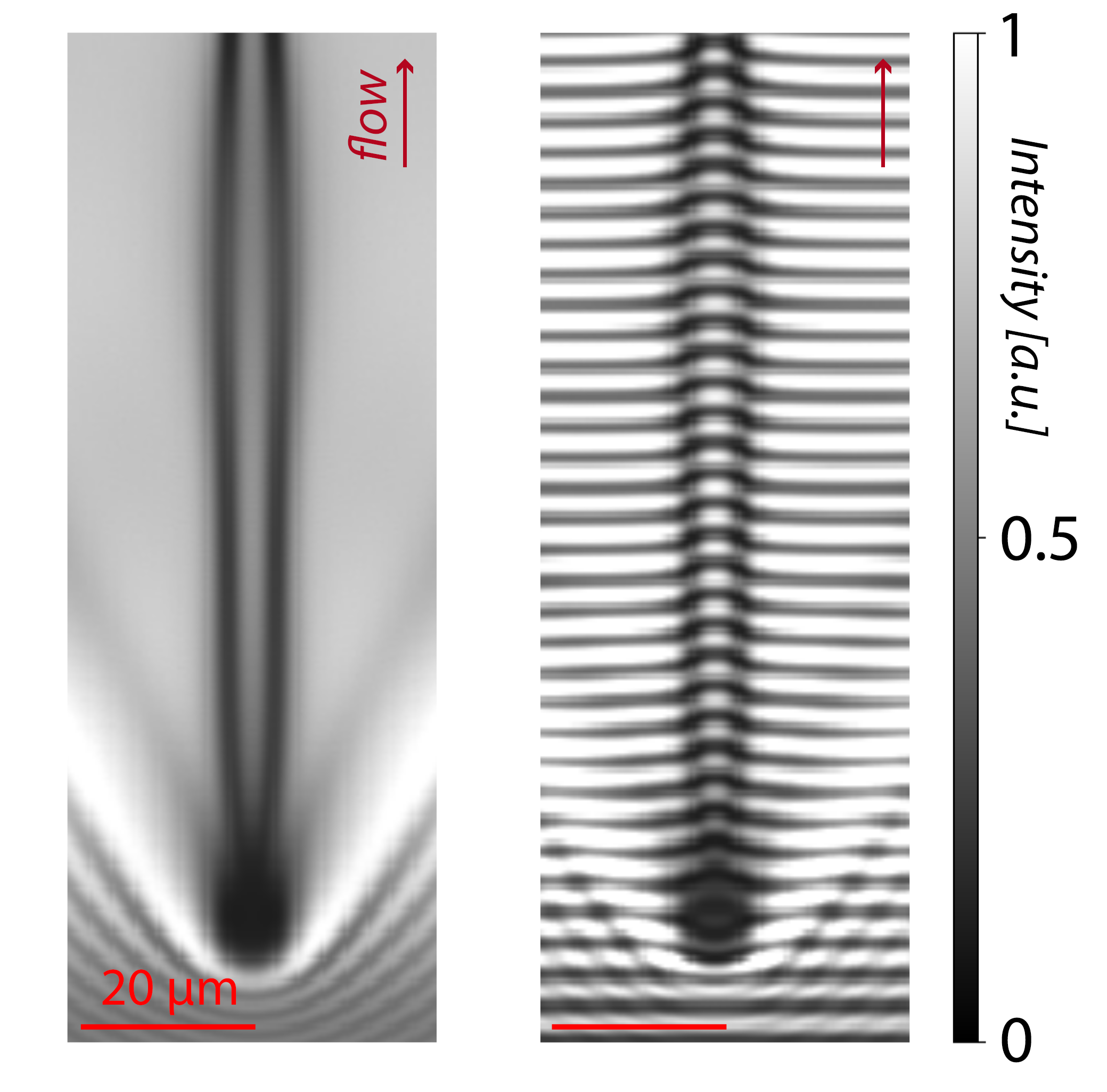}
    \caption{Simulation of the spontaneous generation of dark parallel solitons. Using the driven-dissipative Gross-Pitaevskii equation, intensity (left) and interferogram (right) simulation of a dark soliton pairs spontaneously generated in the wake of a defect. }
    \label{fig:SolParrSimu}
\end{figure}

The results of the numerical simulations are presented in figure \ref{fig:SolParrSimu}. They show an excellent agreement with the experimental data of figure \ref{fig:SolParrExp}. The left image is the polariton density map, switched on to the upper branch of the bistability cycle by combination of both the localized intense seed and the bistable extended support. 
The interference pattern on the right exhibits a clear phase jump all along the propagation, confirming the solitonic nature of the intensity pattern. 
As expected, the solitons are sustained for a macroscopic distance and stay parallel to one another during their propagation.

\section{All-optical imprinting of dark soliton molecules in a polariton superfluid}

We described in the previous section how to use the bistable behaviour of the polariton system to enhance the propagation length of a polariton superfluid, simultaneously getting rid of the phase constraint of the pump. 
This allowed us to observe the spontaneous generation of quantized vortices and dark solitons and their propagation for over a hundred microns. However, their generation was not controlled, since it depends on parameters out of reach; in particular, the necessary presence of a structural defect to induce the turbulence leading to the topological excitations \cite{Lerario2020, Lerario2020a}.

The goal in this section is to show how to overcome this limitation and to be able to generate solitons on demand. 
It is realized by imprinting a phase modulation on the system, leading to the formation of dark solitons that can evolve freely on the nonlinear fluid. 
Such controlled impression of solitons provides many tools for the detailed study of their hydrodynamic behaviour, as the main parameters of the soliton pattern, such as their shape and position, can be tuned at will. 
It results once again in the unexpected binding mechanism between the imprinted solitons, leading to the propagation of a dark soliton molecule. \cite{Maitre2020}.

The main difficulty of this experiment is to combine two regions of the systems which exhibit a different behaviour: namely a region whose phase is imposed by the pump and a free propagation area, where despite to the fact that the phase of the pump beam is flat, yet the system can develop and sustain topological excitations.

\begin{figure}[h]
    \centering
    \includegraphics[width=0.9\linewidth]{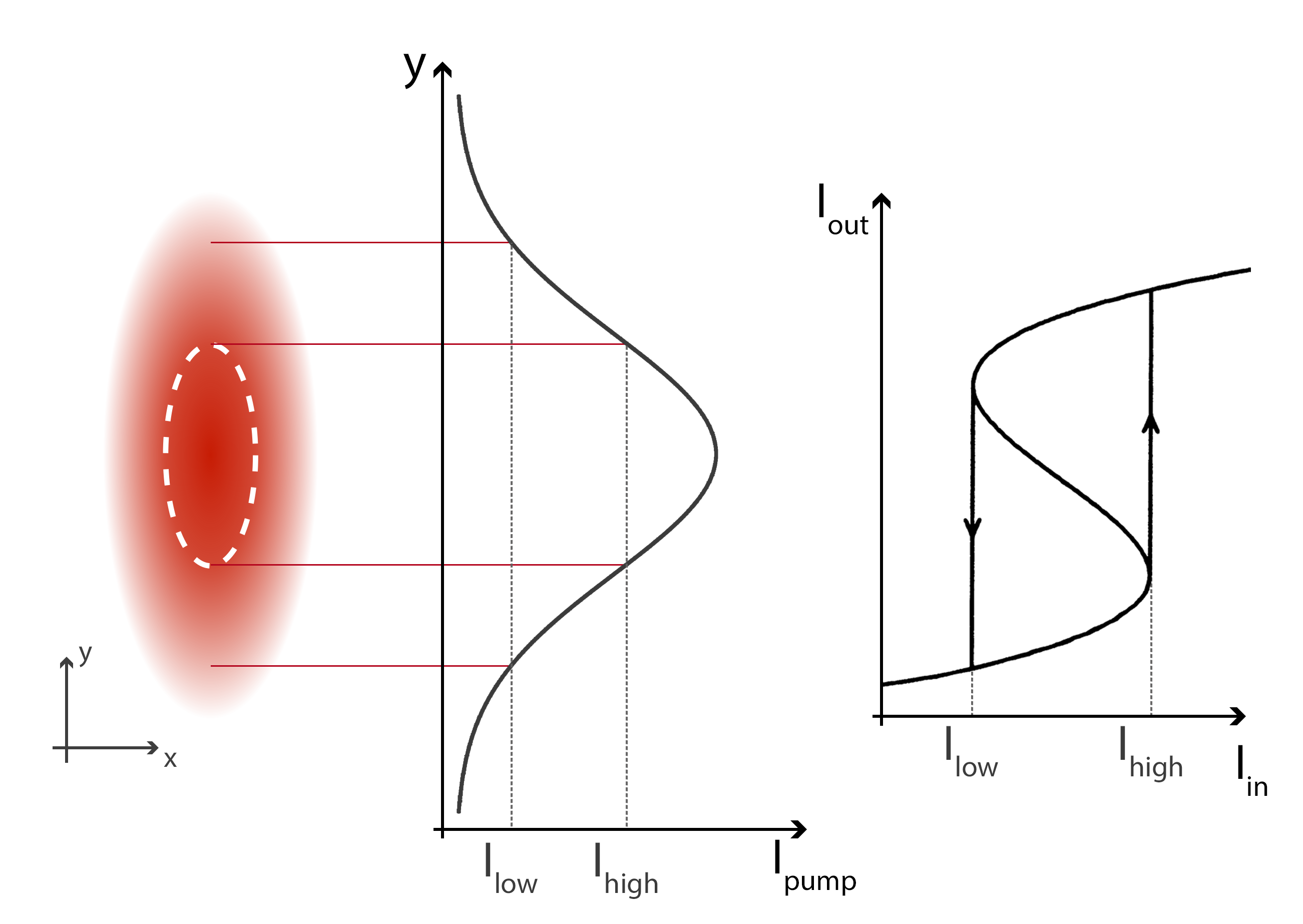}
    \caption{Beam intensity and bistability. Left, spatial distribution of the excitation beam. The white dashed line illustrates the position of I\textsubscript{high}, upper intensity threshold of the bistability cycle, as shown in the intensity profile on the center image. On the right, the theoretical bistability cycle with the definition of I\textsubscript{high} and I\textsubscript{low}}
    \label{fig:BeamInt}
\end{figure}

Once again, it is achieved by using the properties of the optical bistability. Indeed, the input intensities above the bistability cycle impose their phase to the fluid, while this constraint is released for the input intensities within the bistability cycle.
The two situations are obtained by exploiting the gaussian shape of the excitation beam, as illustrated in figure \ref{fig:BeamInt}. 
The spatial distribution of the beam is plotted on the left, where the white dashed line indicates the position of the high intensity threshold of the bistability $I_{high}$ (see right picture). 
All the area inside this circle is above the bistability cycle, as shown in the profile in the center. Therefore, it fixes the phase of the fluid and corresponds to the effective impression region.
Outside of this circle, most of the beam intensities are within the bistability cycle: the phase is not imposed anymore and the system is able to sustain the free propagation of topological excitations.
The beam is elongated in the \textit{y} direction in order to flatten its profile and extend the bistable region where the solitons free propagation will be studied.

As the solitons induce a phase jump on the system, their implementation can be done by modeling the phase of the excitation. To do so, we use a Spatial Light Modulator, a liquid-crystal based device that can shape the wavefront of an incident light beam. 

The phase modulation induced by dark solitons is a phase jump of \textpi: the phase profile corresponding to a pair of dark solitons is thus an elongated region \textpi-shifted compared to the background beam. 

\begin{figure}[h]
    \centering
    \includegraphics[width=0.6\linewidth]{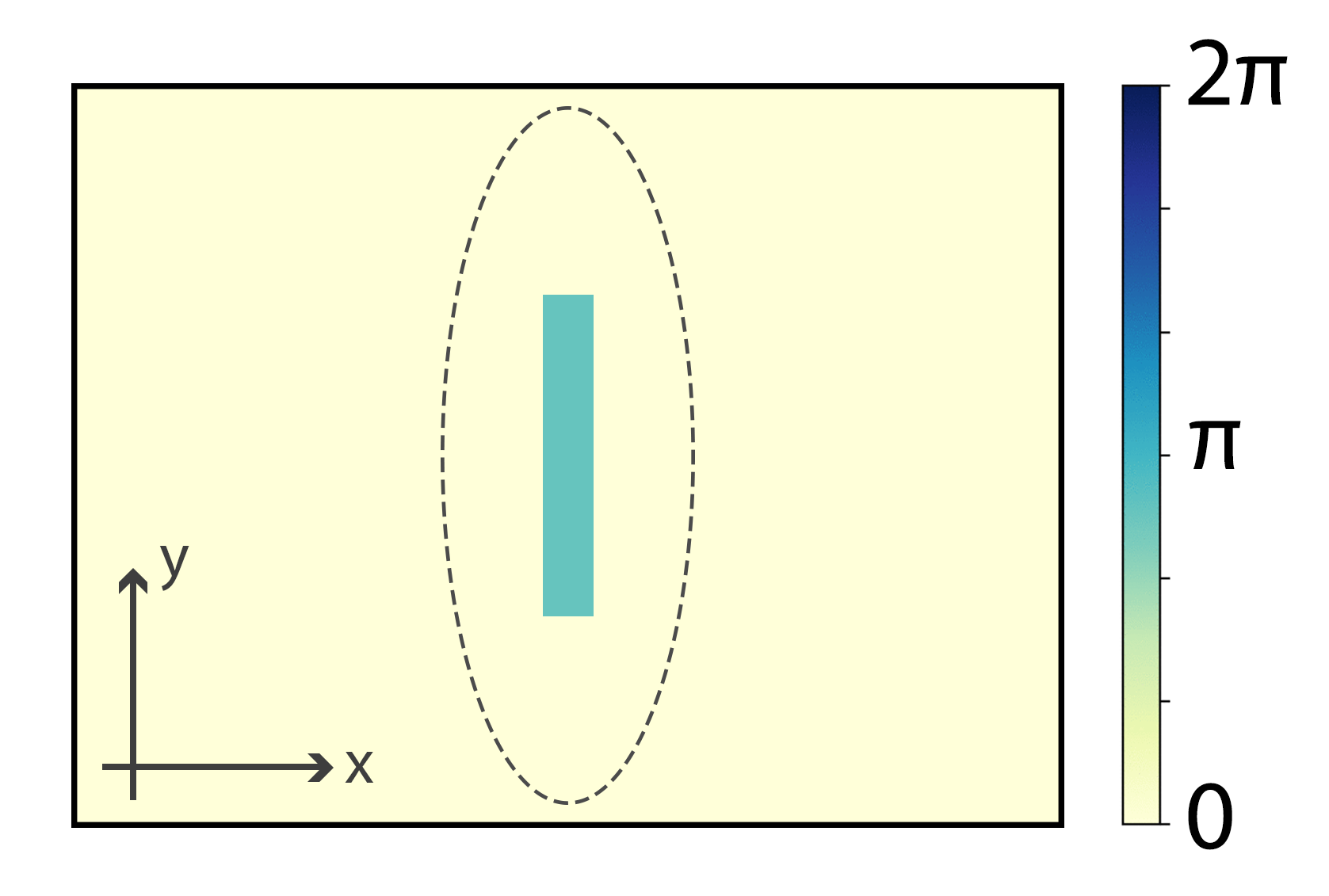}
    \caption{SLM profile. Typical \textpi-shifted rectangular shape generated by the SLM on the excitation beam. Its position and shape are easily tunable. The dashed line delimits the region of the beam which is above the bistability cycle.}
    \label{fig:SLMpatt}
\end{figure}

The figure \ref{fig:SLMpatt} shows a typical phase pattern generated by the SLM on the beam wavefront: it results, as desired, in a rectangular phase jump of \textpi\ which consequently induces an intensity dip. The dashed line delimits the region of the beam which is above the bistability cycle. 

\begin{figure}[h]
    \centering
    \includegraphics[width=\linewidth]{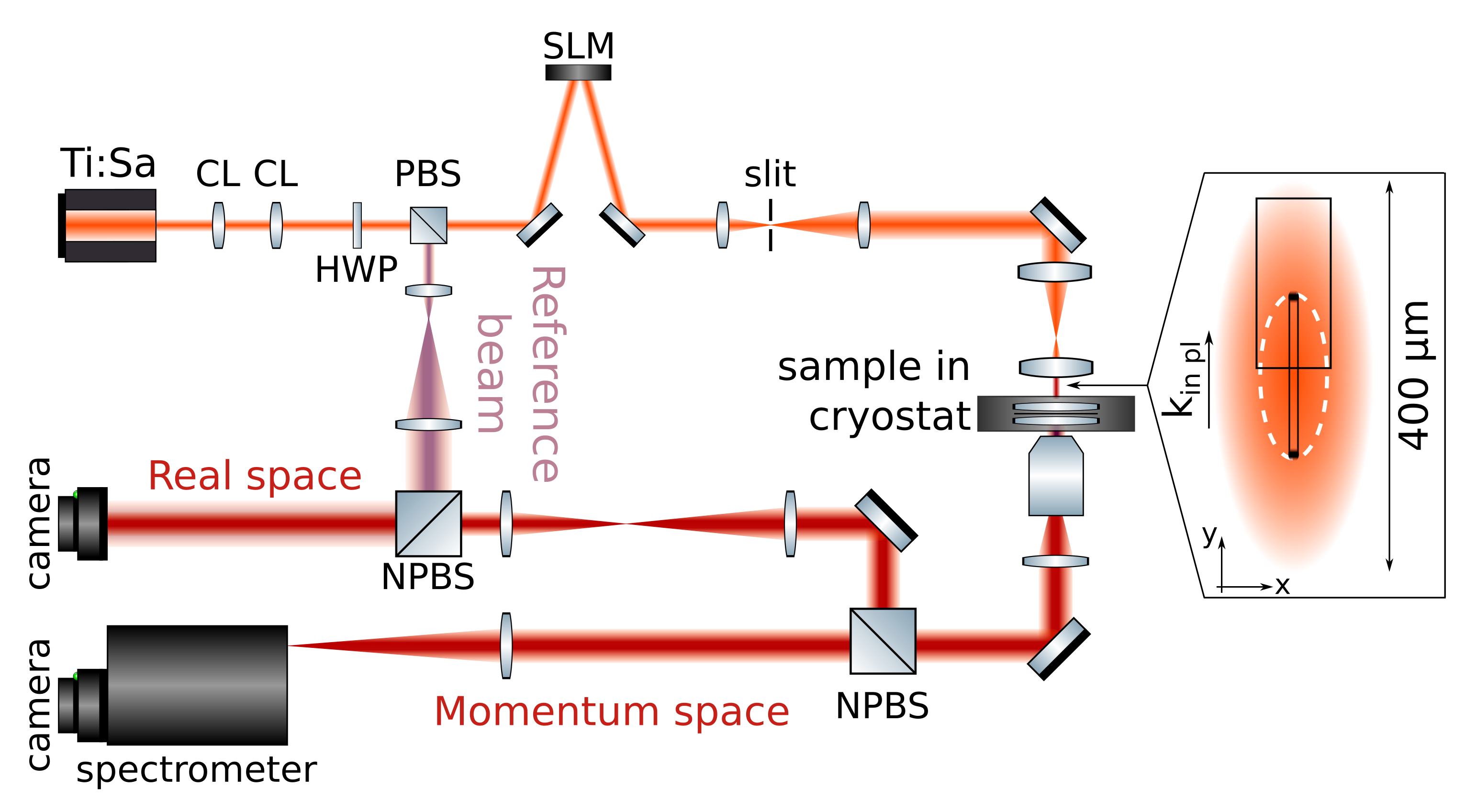}
    \caption{Experimental setup. The excitation beam is designed by the SLM and filtered through the slit. It is collimated on the sample, imaging the SLM plan. The inset illustrates the beam configuration: the solitonic pattern is in the center, where the intensity is above the bistability limit, located at the white dashed line. The detection field of view is delimited by the black rectangle, shifted from the center in order to focus on the bistable region and the solitons free propagation. The detection is done in real and momentum space, so that the experimental conditions can be associated with the corresponding intensity and phase maps.}
    \label{fig:SetupImpr}
\end{figure}

The experimental setup is sketched in figure \ref{fig:SetupImpr}. The laser source is a Titanium Sapphire, and its spot is elongated in the \textit{y} direction by two cylindrical lenses (CL). The beam is then split in two by a polarizing beam splitter (PBS) preceded by a half-wave plate (HWP), which allows for a precise control of the power sent in each arm. 
The phase front of the main beam is shaped by the SLM, then is filtered by the slit to smooth the phase jump in the direction of the flow. This beam is sent collimated to the cavity, so that the phase jumps are well defined onto the sample.
The inset illustrates the excitation beam configuration on the sample: the solitonic pattern is placed in the center, where the intensity is above the bistability limit, delimited by the white dashed line. 
The beam enters the cavity with an appropriate in-plane wave vector that gives an upward flow to the polaritons.
The black rectangle is the detection field of view: it is shifted on top of the illuminated region to observe the solitons free propagation through the bistable area. 

As usual, the detection is done in both real and momentum space. The real space gives the intensity map of the cavity plan, as well as information on the phase pattern through the interference with the reference beam previously separated from the laser beam.
The experimental conditions of the system are extracted from the momentum space images.

\begin{figure}[h]
    \centering
    \includegraphics[width=\linewidth]{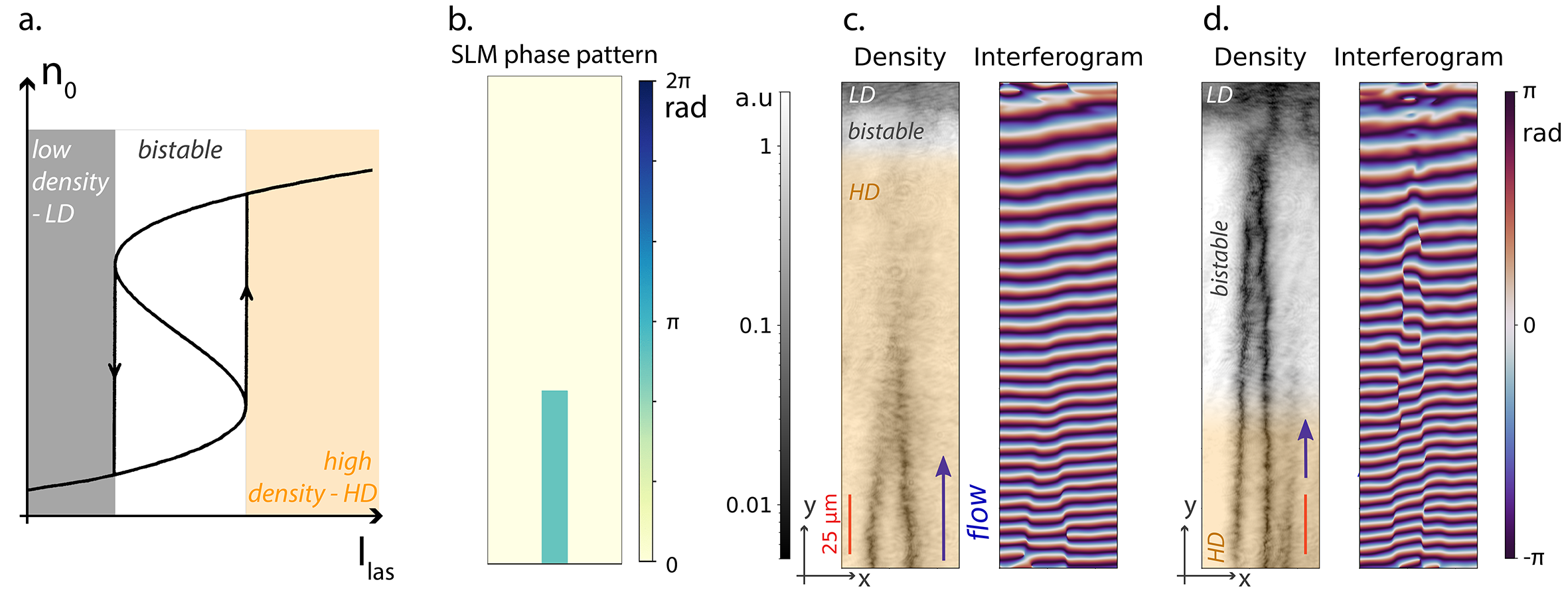}
    \caption{Impression of dark parallel solitons. a. Theoretical bistability profile and the three associated intensity ranges. b. SLM phase pattern with the same field of view as the detection: the phase modulation are only sent on the bottom part of the images.
    c. High power density and phase maps. Almost all the illuminated area is in the HD regime: the phase is fixed by the driving field and replicates the pattern designed by the SLM. d. Same configuration as c. at lower power. The bistable region has extended toward the beam center, and reached the top part of the solitons. The dark solitons propagate through the bistable area, until the low density region where the nonlinear interactions are too low to sustain them.}
    \label{fig:ImprParallel}
\end{figure}

The essential role of the bistability in the propagation of the solitons is explained in figure \ref{fig:ImprParallel}.
The figure a. reminds the S shape of the bistability curve
and the three associated intensity regions: below the cycle, the low density region in grey, denoted as LD; above the cycle, the high density region highlighted in yellow and denoted as HD; and the bistable cycle left blank.

On b. is plotted the phase pattern designed by the SLM, with the same field of view as the detection: we can see that the phase modulation is only present on the bottom part of the images.
Figures c. and d. are realized in the exact same conditions except for the total intensity of the excitation. In c., the total laser power is high, which puts almost all the illuminated area above the bistability cycle: the yellow HD region covers the major part of the picture. In the high density region the properties of the fluid are fixed by the pump: this area is therefore a replica of the driving pump field. 
Indeed, the solitons are imprinted only in the bottom part of the picture, while on top, the phase and the intensity of the pump beam are flat.

Figures d. are obtained from the c. ones by gradually decreasing the input intensity. The bistable region expands toward the center of the beam, and eventually reaches the top part of the imprinted solitons.
The solitons then propagate through the bistable region, even though the region between the solitons is out of phase with the driving field.
Indeed, the dark solitons in d. are clearly visible within the bistable region, inducing a phase jump of \textpi\ all along their propagation. The propagation is sustained as long as the system is in the bistable regime. As the illuminated region is finite, the solitons will reach its border: in the low density region, the nonlinear interaction are too low to sustain dark solitons.

To find the good configuration for the solitons to propagate through the bistable region, several parameters need to be finely tuned. In particular, the total intensity of the pump has an important impact on the soliton propagation, as they need a bistable fluid to be sustained and propagate. 

\begin{figure}[h]
    \centering
    \includegraphics[width=\linewidth]{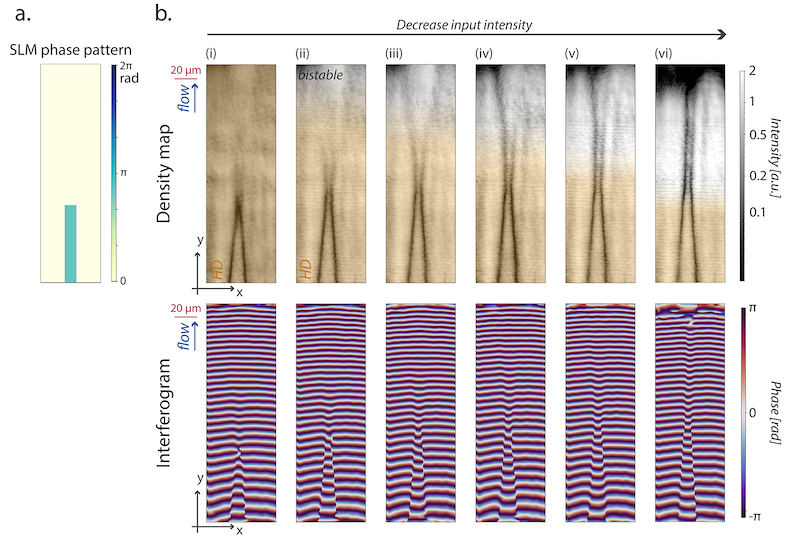}
    \caption{Scan of the input power. a. Phase pattern designed by the SLM with the same field of view as the detection.
    b. Density (top) and phase (bottom) maps for different values of the total input intensity. On (i), the power is maximum and the phase is fixed everywhere. The intensity gradually decreases and with it the size of the HD region in yellow: the solitons propagate and open, until the bistable part reaches the imprinted solitons in (vi) where they align and propagate parallel.}
    \label{fig:ScanInt}
\end{figure}

To qualitatively study the influence of the pump intensity, several images are recorded for different values of the total pump intensity. The results are presented in figure \ref{fig:ScanInt}. The SLM phase pattern is presented on a., again with the same field of view as the detection images, and the experimental images are shown in b.
The top line shows the density maps and the bottom one the interferograms.
The input power is gradually decreased from picture (i) to (vi). The flow is from bottom to top, and the yellow colored regions indicates the area above the bistability cycle.

In picture (i), the total power is high: the fluid is above the bistability cycle on the whole picture. Its phase is therefore fixed, consequently the solitons are imprinted only in the bottom part of the image, where the phase front of the beam is shaped by the SLM.
From picture (ii) to (v), as the power decreases, the bistable region expands. The solitons propagate further and further but the system still can not perfectly sustain them: they are grey as their phase jump is lower than \textpi\, and they open and vanish along the flow. 
The phase maps confirm as well that the phase modulation induced by the solitons vanishes with them.

Finally, in picture (vi), the bistable area joins the top part of the imprinted solitons. They are then able to align to each other, and to remain dark and parallel all along their propagation. In this case, their phase jump is \textpi\ and stays constant.
They are sustained through the whole bistable region, and vanish only at its edge, where the system jumps to the low density regime.

This set of measurement clearly confirms the necessity to be in the bistable regime (namely inside the bistability cycle) to achieve the free propagation of dark solitons in a resonantly pumped polariton fluid.

The imprinting method provides a large flexibility on the shape of the imprinted phase pattern: the SLM pattern is easily tuned and several other configurations can be implemented, in order to study the solitons behaviour. In particular the imprinting technique is fully scalable and allows generating stable pattern of multiple solitons, which opens the way of the study of soliton lattices in polariton superfluids. 

\begin{figure}[h]
    \centering
    \includegraphics[width=0.9\linewidth]{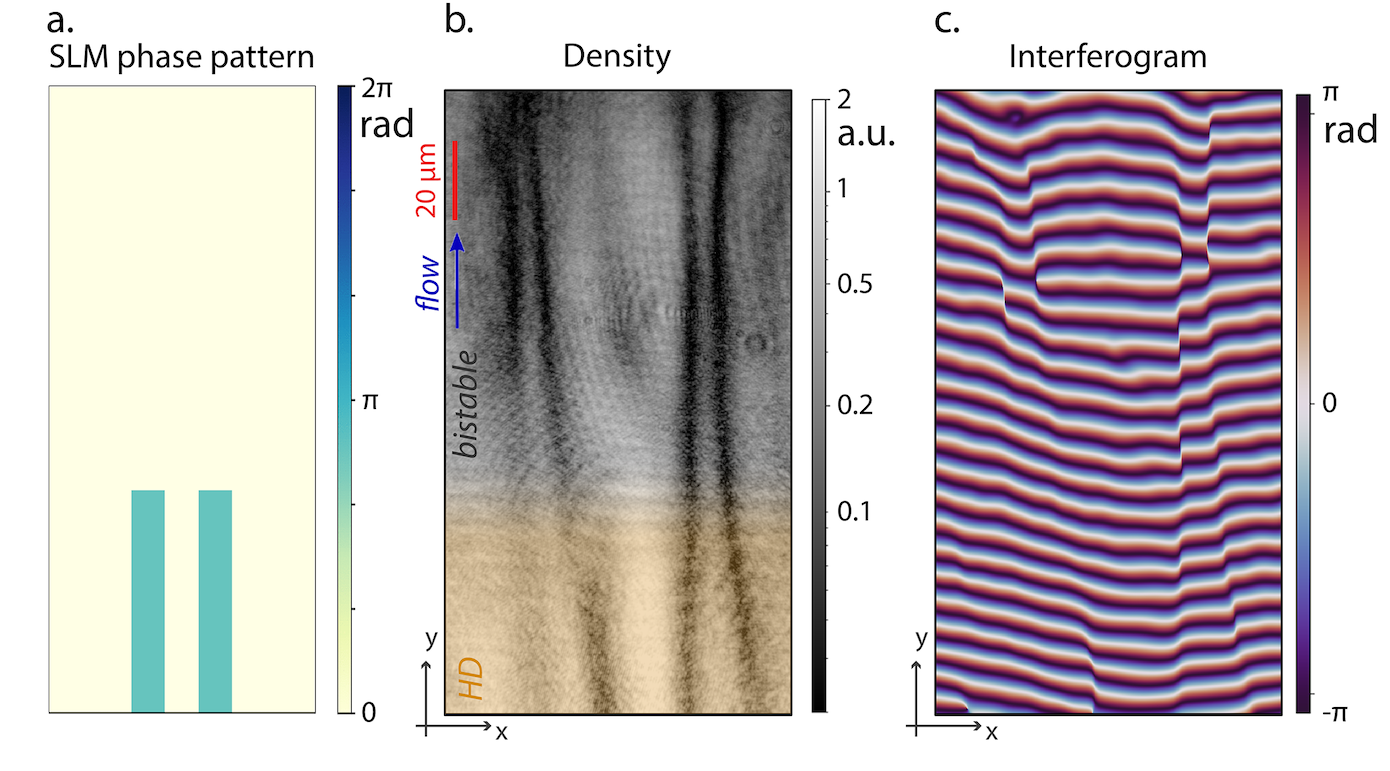}
    \caption{Four solitons imprinting. a. SLM phase pattern corresponding to the detection region. b. Density map of the fluid. The two solitons pairs are imprinted on the yellow region and propagate through the bistable white one. c. Interferogram of the fluid. The phase jump propagates with the solitons.}
    \label{fig:Impr4Sol}
\end{figure}

As an example of the scalability of such method, we show in figure \ref{fig:Impr4Sol} a double pair of solitons.
 On the left, figure a. is a scheme of the SLM phase pattern. 
It represents the top part of the beam, coinciding with the corresponding detection pictures, plotted on figure b. and c. and showing the density map and the interferogram, respectively.

The double pair of solitons is realized by sending two rectangular shapes in phase opposition with the background thanks to the SLM. The imprinted phase pattern is designed so that each of the four solitons is equidistant from its neighbor (yellow part of figure \ref{fig:Impr4Sol}.b.). During their free propagation in the bistable region of the fluid (grey part of figure \ref{fig:Impr4Sol}.b.), they get closer to their respective pair, so that the area in phase with the driving expands while the one in phase opposition is reduced. 

\section{Snake instabilities, soliton breaking and vortex streets}
In the experiments described in the previous section we decided to generate and study stable solitons and for this purpose we deliberately chose to work in a supersonic regime with high speed polariton flows \cite{Kamchatnov2008a}. However, the flexibility of the imprinting technique and the full control of the fluid
velocity that can be achieved, easily allow one to imprint dark solitons on
polariton fluids with deeply subsonic velocities. In these
conditions, it is well known that dark solitons are unstable against the snake instabilities and break into quantum vortex-antivortex
pairs, a behaviour which is a quantum analog of the
classical von Kármán vortex street \cite{kwon2016observation}. Therefore the imprinting technique can be exploited for the systematic study of quantum turbulence.

As an example of such possibilities, in this section we investigate a new configuration to generate solitonic pattern within a static polariton fluid. 
It uses a transverse confinement within an intensity channel to create a pair of dark solitons, which decays into vortex streets due to the disorder of the system. We observe the soliton snake instability leading to the formation of symmetric arrays of vortex streets, which are frozen by the pump-induced confining potential allowing their direct observation and a quantitative study of the onset of the instabilities.

The setup we used for this experiments is quite similar to the one of the soliton impression, presented in the previous section.
Two main differences have yet to be noticed.
First of all, to facilitate the development of the instabilities, the experiment needs a static fluid: the excitation is therefore sent at normal incidence, and the in-plane wavevector is zero.
Then, the 1D elongated channels which confine the dark solitons are created by shaping the intensity of the excitation beam, while its phase is not modulated anymore.   

Indeed, although the Spatial Light Modulator (SLM) usually shapes the phase front of the incoming light beam, we can also use it as a intensity modulator, and thus design at will an intensity pattern on the beam.
To do so, a grating with a controlled contrast is imprinted in a specific region of the pumping beam by the SLM.
When the full range of the SLM grey scale is used for the grating impression, all the light is diffracted to the  grating first order.
However, by using  a fraction of the scale, only a portion of light is diffracted, resulting in a darker region dug inside the non diffracted zero order beam.
This way, by tuning the grating contrast of the SLM screen and filtering out the first order in the Fourier plane, we can shape a beam with a designed intensity pattern. Again, as it is controlled via a computer, the shape is easily tunable.

The detection is realized in both real and momentum space. The excitation conditions are extracted from the momentum images, while the real space gives us the density map as well as the phase from interferences with a reference beam.

\begin{figure}[htbp]
\centering
  \includegraphics[width = 1\linewidth]{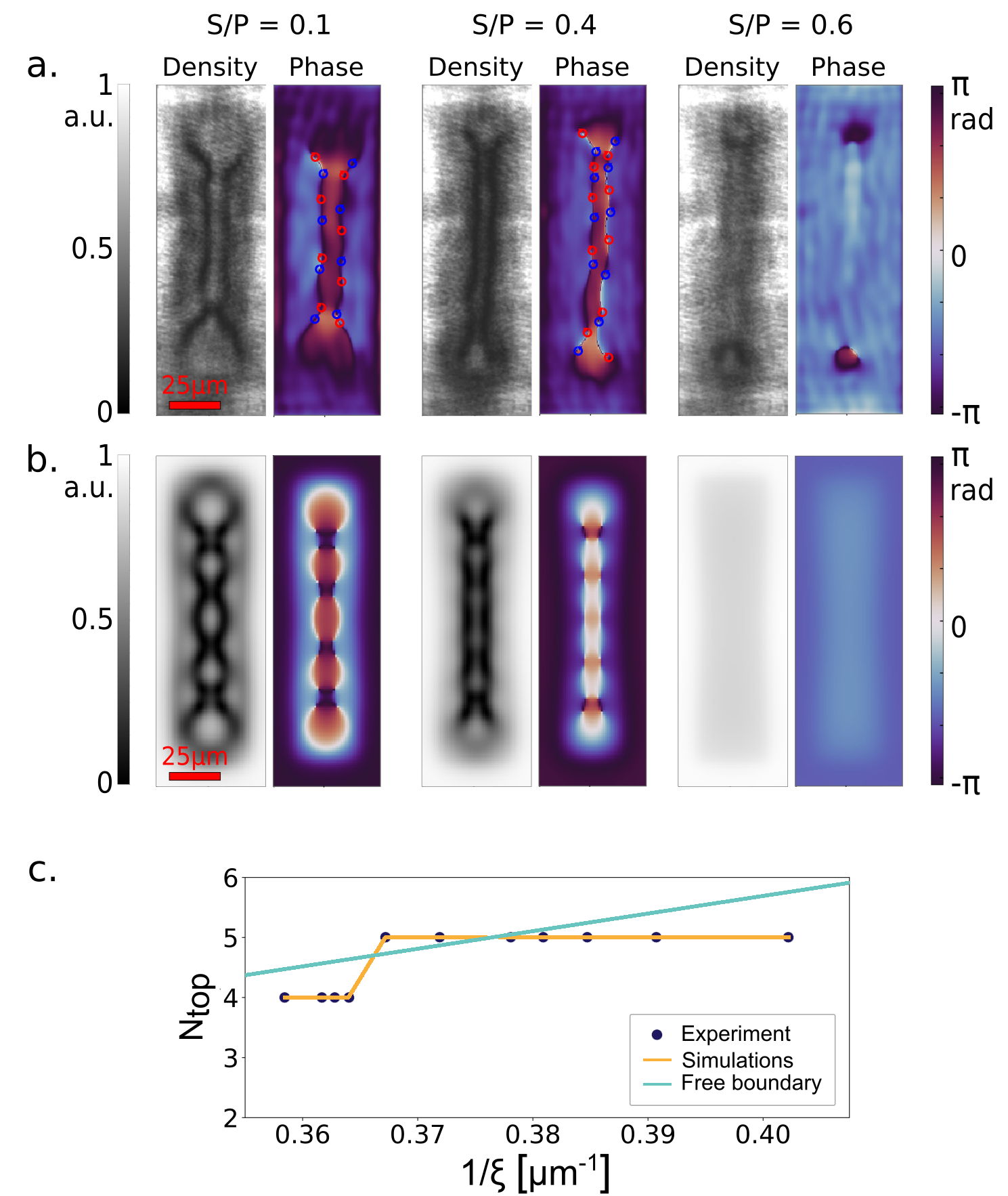}
\caption{\label{fig:fig3} {Panel (a): experimental fluid density and phase} maps for a channel with length and width  respectively l = 150 $\mu$m, L = 23 $\mu$m; the ratio S/P increases from left to right.
{Panel (b)}: corresponding numerical simulations. {Panel (c)}: Evolution of the number of VA pairs N$_{top}$ in the channel for different values of $1/\xi$. The black dots are the experimental data. The yellow curve gives the simulations results under the same experimental conditions whereas the blue curve shows the vortex density for an infinite channel.}
\end{figure}

We consider a channel with both the ends closed by high-density walls (Figure \ref{fig:fig3}). The channel width of 23 $\mu$m is chosen in order to generate a single dark soliton pair, which evolves toward a stationary frozen vortex street due to the snake instability.

In figure~\ref{fig:fig3}, the panel {(a)} shows the measured intensity and phase distributions in the channel for increasing ratios of the intensity inside and outside the channel, called $S$ and $P$ respectively.  The panels {(b)} shows the corresponding numerical simulations obtained by solving the system of coupled equations. A symmetric array of vortex-antivortex (VA) pairs, similar to a Von-Karmann vortex street is visible for $S/P = 0.1$ and $0.4$, respectively. 
A soliton pair is indeed unstable in this regime \cite{Koniakhin2019a} against modulational "snake" instability. It therefore breaks into the observed VA chain. In free space, these chains would dynamically evolve \cite{Dutton2001} to eventually disappear. Remarkably, the presence of the confining potential allows to freeze the snake structures at a given stage of their evolution and to easily observe them in a steady state CW experiment.

The particle density in the channel is associated with the healing length of the fluid $\xi$ which sets the dark soliton width and the vortex core size. 

It also naturally sets the spatial period at which the instability develops along the main channel axis \cite{Koniakhin2019a} and therefore the number of VA pairs which appear in a channel of a finite length $L$. The panel (c) of figure~\ref{fig:fig3} shows the number of pairs experimentally observed versus $1/\xi$ and in red, the theoretical value obtained by numerically solving the system of the coupled equations, which are in excellent agreement. They both show a step-like increase due to the quantization imposed by the finite channel length. The blue line shows the expected number of vortices per length $L$ in an infinite channel which is proportional to $\sim 1/\xi$.

\section{Conclusions}
In this article we have reviewed our recent results on the polariton optical bistability and its related properties.  

By implementing the theoretical proposal \cite{Pigeon2017} we demonstrated the possibility to use bistability exhibited by the polariton system under resonant pumping together with a pump-support configuration to  sustain  topological excitations such as quantized vortices and dark solitons \cite{Lerario2020, Lerario2020a} over hundreds of microns, greatly enhancing their propagation length compared to the previous observations \cite{Amo2011}.

This experiment also revealed a very unexpected behaviour of the dark solitons: the presence of the driving field imposed by the bistable pump compensates the dark solitons repulsion as usually observed in an undriven system \cite{Amo2011, Hivet2012a}. 
In this new configuration, dark solitons align to each other and propagate parallel.
Moreover, to achieve a full control on the formation of the topological excitations we developed a new all optical imprinting method by accurately shaping the excitation beam with a Spatial Light Modulator, and we managed to generate on demand dark solitons on a polariton fluid \cite{Maitre2020}.
Once again, due to the presence of the driving field, we observed this binding between the solitons, propagating parallel as a dark soliton molecule.

The flexibility of the imprinting method opens the way to deeper studies of quantum turbulence phenomena. In particular we generated solitonic structures in guided low-density channel in a static polariton fluid \cite{Koniakhin2019a} and observed their breaking into vortex streets due to transverse snake instabilities \cite{Claude2020}.

The imprinting technique could give significant contributions to elucidate open questions as the formation of turbulent cascades in the presence of dissipation at all length scales \cite{koniakhin20202d}, as in polaritons, and the validity of the entropy arguments that explain the formation of inverse cascades in equilibrium two-dimensional systems \cite{simula2014emergence,gauthier2019giant,johnstone2019evolution}.

\acknowledgments
This work has received funding from the French ANR grants ("C-FLigHT" 138678 and "Quantum Fluids of Light",  ANR-16-CE30-0021), from the ANR program "Investissements d'Avenir" through the IDEX-ISITE initiative 16-IDEX-0001 (CAP 20-25), from the European Union Horizon 2020 research and innovation programme under grant agreement No 820392 (PhoQuS).
QG, AB, and DS thank the Institut Universitaire de France (IUF) for support. SVK and DDS acknowledge the support from the Ministry of Education and Science of the Russian Federation (0791-2020-0006).

\bibliographystyle{ieeetr}
\bibliography{LKB-bibs-bib_thesis.bib}

\end{document}